\documentclass[authoryear,preprint,12pt]{elsarticle}

\usepackage{graphicx}

\usepackage{amssymb}

\usepackage{enumitem}

\usepackage{rotating}

 \usepackage{lineno}

\journal{Planetary and Space Science}

\begin{document}

\begin{frontmatter}



\title{A confidence index for forecasting of meteor showers}


\author{Jeremie Vaubaillon}

\address{IMCCE, Paris Observatory, PSL, 77 Av Denfert Rochereau, 75014 Paris, France}

\begin{abstract}
The forecasting of meteor showers is currently very good at predicting the timing of meteor outbursts, but still needs further work regarding the level of a given shower.
Moreover, uncertainties are rarely provided, leaving the end user (scientist, space agency or the public) with no way to evaluate how much the prediction is trustworthy.
A confidence index for the forecasting of meteor showers is presented. 
It allows one to better understand how a specific forecasting has been performed.
In particular, it underlines the role of our current knowledge of the parent body, its past orbit and past activity.
The role of close encounters with planets for the time period considered is quantified as well.
This confidence index is a first step towards better constrained forecasting of future meteor showers.
\end{abstract}


\end{frontmatter}


\section{Introduction} \label{sec:intro}

The prediction of meteor showers on Earth has been the topic of much research since the XIXth century.
The observation of recurrent outbursts (such as e.g. the Leonids every 33 years more or less) has been the first motivation to conjecture about future events.
In addition, the link between meteor showers and comets was established by Schiaparelli \citep{Romig1966} and shortly later the first forecastings were based on the orbit of the parent comet.
One famous failure was however the expected return of the Leonids in 1899, as well as in the three following perihelion returns of comet 55\/P.
It was not %
 before \cite{KondratevaReznikov85} and later on \cite{McNaughtAsher1999}  that an estimate of the time of a shower outburst was correctly predicted.

If the timing of meteor showers is currently well constrained by todays works, the level of the shower still poses a challenge to astronomers.
Failures at predicting a correct level of a shower has consequences for researchers, space agencies and the public.
Beside the disappointment aspect of missing on an opportunity which might end up being a waste of time, protection procedures for spacecraft require lots of time and energy.

The success of predicting a shower was enabled by understanding that meteoroids and comets have similar yet independent orbits and orbit evolution.
Today methods are more or less all the same and are based on the propagation of the orbit of test particles released from the parent body, from the time of ejection until it passes near the Earth.
Refinements include: ejection over an arc of the orbit, massive simulation of test particles, update of the ejection velocity %
 (i.e. taking into account the physics behind the ejection process) .
Among the authors performing such forecasting, we find \cite{McNaughtAsher1999,2000EM&P...82..149L,Vaubaillon.et.al2005,2008EM&P..102..111W}.
However, apart from those, no new method has been developed recently.

Surprisingly, in spite of the quality of the work dedicated to meteor shower forecasting, no uncertainty has ever been published to my knowledge.
The first reason probably comes from the dynamical approach of the forecasting, which was the Achilles heel until 1999, and the focus of many works.
However, seventeen years later this has not improved.
The second reason most probably comes from our ignorance in so many physical quantities of the parent body as well as its past dynamical behavior.

The difficulty of providing uncertainties can certainly be overcome, by going through a rigorous analysis of every step leading to a given forecast.
However, one might argue that such a refinement might not tell us much, again because of our uncertainty in e.g. the parent body parameters.
In other words, it might be hard to define a credible uncertainty of a physical quantity for which even orders of magnitude cannot be estimated.

Because the end user of the forecasting still needs a way to know how much (s)he can trust a given prediction, this paper presents a different approach.
The idea is to provide the scientists, space agencies and amateurs some knowledge regarding the circumstances under which the predictions were performed, and inform them regarding the chances of success, especially in terms of the level of the shower.
I hope that by doing so every reader of future forecasting can have a proper idea of how much (s)he can trust the forecasting.

The paper first presents in section \ref{sec:strat} a reflexion on the way meteor shower predictions are performed today and underlines the location of greatest uncertainties.
Then in section \ref{sec:confidx} a confidence index is presented that provides the end users with enough information to have an idea of how much one can trust the forecasting.
Last but not least in section \ref{sec:appli} some confidence indices are listed for past and future showers.

\section{Strategy} \label{sec:strat}

In order to perform the forecasting of the timing (T) of a meteor shower, one needs to know :
\begin{enumerate}[label=T\arabic*]
\item \label{PB} the parent body 
\item \label{pastorb} the past orbit of the parent body
\item \label{eject} how meteoroids are ejected from the parent body
\item \label{evolv} how meteoroids orbits evolve in the Solar System.
\end{enumerate}

In order to perform the forecasting of the level (L) of a meteor shower, one needs to know %
 \ref{actShw} or \ref{actPB} as well as \ref{conv}, as explained below :
\begin{enumerate}[label=L\arabic*]
\item \label{actShw} the past activity of the shower
\item \label{actPB} the past activity of the parent body
\item \label{conv} a way to convert this activity into a ZHR.
\end{enumerate}

Point \ref{evolv} is quite well understood today, and point \ref{eject} does not matter much, since the knowledge of an order of magnitude is good enough to perform a correct prediction.
The reason is that anyway meteoroids are ejected with a distribution of velocities and a distribution of heliocentric distances.
The identification of a parent body has recently seen a huge development thanks to multi-years surveys \citep{2011Icar..216...40J,2015P&SS..118...38R,2014pim4.conf...34C}.
The accumulation of tens of thousands of meteoroid orbits allows one to better recognize otherwise undetected showers, and dynamical links are based on orbital similarity.
In a similar way, the discovery of new thousands of NEOs makes it more likely to find a parent body for a given new shower.
In other words, point \ref{PB} is being currently revolutionized by huge amounts of data and data mining.
Similarly, point \ref{actShw} is being currently refined for the same reasons.
However, if the basic knowledge of the activity of a shower is poorly constrained, needless to say that any estimate of a future shower cannot be accurate.
This is particularly preventing the performance of predictions on other planets as Earth (Mars and Venus being the currently most wanted ones).
 Point \ref{conv} is usually straightforward by converting a 3D particle density into a 2D density, or by comparing the past encounter circumstances (e.g. distance between the center of a trail with the path of the Earth) with the forecasted ones \cite{McNaughtAsher1999}. 

What is left are points \ref{pastorb} and \ref{actPB}, forming the source of most uncertainties, in my opinion.
The past orbit of famous parent bodies (such as 1P/Halley, 109P/Swift-Tuttle) can be useful by telling us that their orbit is stable enough and that their activity spanned several centuries.
However, this might not directly explain the level today of e.g. the Orionids and Perseids if the encountered particles are older than the oldest record of the comet.
This is unfortunately indeed the case for 1P and 109P, and the reason why the predictions of the Perseids are mainly performed by the International Meteor Organization and based on past observations of the shower \ref{actShw}, provided it is stable enough.

In most cases, the past orbit of a parent body is problematic, by lack of past observations.
Even if one can dig in historic records, one cannot find anything beyond 5000 years ago, which might not be enough for long period bodies \citep{2016A&A...589A.100N}.
Fortunately, as long as the orbit of the parent body is stable enough (see comment below regarding this notion), and its cometary activity either non existent or constant from one passage to another, it is easy to find its past returns, yielding to the forecasting of future showers.
However, usually the past activity is even less constrained than the orbit of the parent body.

Another problem is the stability of the orbit of the parent body.
Even if its orbit today is well constrained, close encounters are prone to dissipate any hope to know its orbit past a certain date.
One famous example is comet 67P/Churyumov-Gerasimenko \citep{2015A&A...579A..78M}, for which it is hard to clearly know its orbit before the 1950s.

Are we therefore doomed in our ignorance of so many important parameters?
Several works tend to provide constraints on the origins of meteor showers, which by such enables to better perform the predictions of future events.
However this is not always feasible.

In this paper, the approach first considers that in complement to all these research works, it is useful to provide information regarding the way predictions are performed, in order to sense the difficulty and uncertainties considered in a given prediction.
The idea is to consider each main source of uncertainty and either label or quantify it.
The confidence index is therefore a code providing information on how the ephemeris of a given meteor shower was calculated.

\section{The confidence index} \label{sec:confidx}

The confidence index is built as a succession of letters and numbers, each having its own meaning and dealing with a specific challenge to perform an accurate forecasting.

\subsection{First letter: the trail index} \label{sec:first}

The first consideration deals with the number of trails the forecasting process is dealing with.
In the most usual and simple case, one trail encountered by the Earth results in a single prediction.
In such a case, the trail index contributing to the confidence index is set to "S" (as in Single trail).

However such a method is unable to e.g. predict the usual background level of the Perseids, as it consists of the superposition of very old trails %
 ($ > 10 k$ years old),  for which the exact origin is unknown.
The simulation of such many trails, providing global information of the shower is feasible but needs to be documented to allow the end user to be warned that the exact origin of the trails is not accurately known (beside the knowledge of the parent body).
In such a case, the trail index is set to "G", meaning that the Global level of the shower was computed.
The end user can therefore quickly know by examining the first letter that a "G" will a priori provide a less accurate prediction than an "S".
Put it in another way, a "G" means that the background of the shower is forecasted, rather than an outburst.
This is of particular use for e.g. the Leonids, known to present rare exceptional outbursts and a low activity otherwise (15/hr).

\subsection{Second letter: year index} \label{sec:second}

The second consideration deals with the uniqueness of the time period for which the prediction is performed.
Most of the time, meteor shower forecasts are computed by considering the particles approaching the planet during a short time period (usually of a few days \citep{1998Icar..133...36B,2005A&A...439..761V}.
Most of the time a given trail is not perturbed enough to present more than one encounter with the Earth for a given year.
In such cases, the "year index", contributing to the confidence index is set to "Y" (as in Year), meaning that the prediction is valid for a given year and includes only the particles crossing the planet at this time.

Now, in the case of a low level shower and even by considering several tens of thousands of particles in the simulations, there might not be enough test particles to compute a level that really makes any physical sense.
One solution is to greatly increase the number of simulated particles \citep{2008EM&P..102..157J}.
However another solution is possible.
In such a case, the idea is to concatenate the contribution of all the particles encountering the planet over several years.
This provides us with an idea of the background activity of the shower, and the location of the stream, rather than the individual location of several given trails.
Such an approach is useful also for parent bodies for which the orbit is not well constrained.
Note that in order to derive a correct timing of the background activity of the shower by following this method, the location of the planet still has to be computed for a short period of time (e.g. several days) and should of course not be concatenated over several years.
By doing so, the change of timing from one year to another can be computed.
In such a case, the "year index" is set to "B", as in "Background".

\subsection{Third element: observation index} \label{sec:third}

The third element of the index deals with points \ref{pastorb} and \ref{actPB}.
It is a measure of the number of observed perihelion passages, versus the number of simulated passages.
It also provides us with information regarding our knowledge of the activity of the parent body.
Indeed, an observation of a return indicates not only the location of the comet (or asteroid), but also provides us with information regarding its activity.
Of particular interest are changes of activity, following either an outburst \citep{2010Icar..208..276R}, a breakup \citep{2010AJ....139.1491V,2011ApJ...740L..11I} or the end of a comet activity \citep{2007AJ....134.1037J}.
However the total absence of observation leaves us with any possible scenario, unless the parent body is observed again at a subsequent passage.
Given the current and future sky surveys, the task of meteor shower forecasting is usually made easy for recent passages.
What is really preventing us to progress is the absence of past observations.
Pre-discoveries are still possible thanks to past surveys \citep{2011AJ....142...28J} or data mining ancient archives \citep{2006pimo.conf...50N}.
Most of the time however it is extremely hard to accurately constrain the orbit and/or the activity of a parent body.
Because of the huge influence of such parameters on the forecasting of meteor showers, the least we can do is to tell the user if a given result comes from an observed return, or if it simply comes from numerical integration of the orbit of the body, considering its activity was constant.
A good illustration of this paragraph is 17P/Holmes \footnote{although it does not create any meteor shower at Earth, we take this example to illustrate the use of observed passages}.
It was discovered in 1892, observed for the following two returns and then lost for nearly 60 years (7 returns) before it was recovered.
The re-discovery allows us to constrain its orbit, and put a limit on its activity.
In 2007 the comet underwent a huge outburst, which completely changed its activity profile.
The 1892 discovery was most likely enabled by a similar outburst to 2007.
To my knowledge, such a change of activity is rarely considered in modern forecasting of meteor showers.
On the other hand, meteor outbursts can help constrain the past activity of a comet \citep{2008EM&P..102..111W}.

The third element of the confidence index is composed of the letter "O" (as in Observations), followed by the number of observed passages $no$ versus the number of simulated passages $ns$.
For example for 17P, one might indicate: $O  6 /13$ to indicate that all the 13 returns of the XXth century were taken into consideration, but only 6 were actually observed.
Most of the time, we have $no < ns$, since it is easy to simulate orbits over several centuries.
However the question arises as to the physical meaning of any long term simulations in the absence of any data to check the results.
If a thorough check is either not feasible by lack of data (as is usually the case), at least the user is informed.
In the extreme case one might consider a newly discovered object for which $no=1$, for which taking $ns=100$ would not make much sense, unless the orbit is very stable and assuming that any past ejection process did not produce significant non-gravitational forces.
Needless to say that such a work would produce highly uncertain results, which might be hard to quantify, but again, at least the user is informed of the way the forecasting was performed.

\subsection{Fourth element: close encounter index} \label{sec:fourth}

The fourth element of the confidence index is a natural following of what was previously mentioned, and deals with the role of close encounters and orbit stability of the parent body (points \ref{pastorb}).
It is worth mentioning that the $f_{M}$ factor introduced by \cite{McNaughtAsher1999} also represents a way to quantify the role of close encounters.
 This quantity is often provided for the given part of the trail that is of interest for a given prediction.
In principle, it can also be computed for the whole trail and by such also measures the effect of time on the spread of the meteoroids within a given trail.  
Here the idea is to provide the user with an idea of how much the orbit of the parent body, at the time of ejection of a given trail, can be trusted, as well as, in a lesser way, its consequences for the trail at the time of the predicted shower.
The idea is to compute a "close encounter index" (CE) by summing all the contributions of all close encounters with the planets (the major perturber being Jupiter), for the duration of the considered simulation.
In practice we have $CE=\Sigma_{t_{min}}^{t_{max}} M_{pla} / M_{sun}  \; 1 / (d V^2)  $, with: $t_{min}$ the time of the ejection of the trail, $t_{max}$ the time of the considered shower, $M_{pla}$ the mass of the encountered planet,  $M_{sun}$ the mass of the Sun,   $d$ the minimum distance of the encounter and $V$ the relative velocity with the planet at the closest distance.
 This expression is inspired from \cite{Valsecchi2003} providing the angle of deflection caused by a close encounter $\tan \gamma/2= \frac{M_{pla}}{dV^2}$. The unit of $CE$ is therefore $s^{2} m^{-3}$.  
In the extreme case, one might have $CE=0.00 \, s^{2} m^{-3}$ if the parent body had no close encounter for the time period considered.
This clearly indicates that the orbit of the parent body does not suffer sudden and drastic changes.
On the other hand, if one gets e.g. $CE=1.0E+04 \, s^{2} m^{-3}$, this indicates that there are numerous close encounters highly changing the orbit of the parent body.
As a consequence, the user can immediately know that such forecasting with such a high close encounter index is a priori much more uncertain than if $CE=0 \, s^{2} m^{-3}$.

However a bit of caution is necessary at this point.
Because the orbit of the parent and the meteoroids are independent, the consequences for a given trail are not necessary immediate and definitely not the same as for the parent, sometimes leading to OMSs (orphan meteoroid streams) \citep{2006MNRAS.370.1841V}.
A high CE value still indicates that at least one giant planet (usually Jupiter) is crossing the stream.
One has to use the CE index in conjunction with the other parameters, especially the observation index.
As an extreme example let us suppose that a new Jupiter family comet has recently been discovered, and found to be the parent body of a weak meteor shower.
In order to produce the forecasting of future meteor showers, the past orbit of the comet is computed over 100 years, i.e. $\sim $ 20 returns.
We have thus an observation index: "O1/20".
Let us suppose now that, in the past, many close encounters with Jupiter happened, increasing the close encounter index to $CE=200 \, s^{2} m^{-3}$.
The combination of those two indices warns the user that the predictions are highly unreliable, mainly by lack of past observations that would have otherwise greatly constrained the orbit of the comet.
Let us now suppose that, on the contrary, the comet was well observed, but still crosses the orbit of Jupiter (as it is usually the case with JFCs).
We might end up with "O16/20" and "$CE=200$", but in this case the user can better trust the predictions, since the orbit and activity of the comet are well constrained for most of the considered period.

 The problem in such a case is that it is hard to disentangle the observation index and the close encounter index, and to know which has a greater contribution than the other.
In a numerical simulation performed to forecast the meteor showers, the effect of close encounters on the parent body as well as on each particle is computed at each time step.
In a sense, we can therefore conclude that all close encounters effects are taken into account thanks to the numerical simulations.
In order to build a confidence index, the close encounter index should therefore indicate the (cumulative) role of the close encounters that escapes our knowledge.
This is why I have chosen to nullify the close encounter index for the time period comprising between the first and last observation of the parent body.
By doing so, $CE$ now indeed reflects our ignorance of the effects of close encounters.
We therefore have: 
$CE=\Sigma_{t_{min}}^{t_{max}} M_{pla} / M_{sun}  \; 1 / (d V^2)  $, with: $t_{min}$ and $t_{max}$ the time period before/after the first/last observation of the parent body.

One last case is possible and has to be discussed here.
Let us suppose that a given prediction was performed by considering several trails (trail index: "G") or considering the contribution of many different years (year index set to "B").
The close encounter index is still computed as the contribution of all the encounters of all the trails and increases rapidly.
In this case, in order to warn the user that the sum was performed on may trails or many years, the prefix "CE" of the close encounter index is changed to "CU" (as in "Cumulative").

\subsection{Summarizing quality label} \label{sec:sum}

All the above mentioned indices provide the end user with information regarding the way the forecasting was performed.
However the end user might not really fully understand nor even care about all these details.
As a consequence, a summarizing quality label is computed.

How to compute such a summarizing quality label?
The presence or absence of observations of the parent body usually makes a huge difference in the confidence one can have in forecasting.
Similarly, the concatenation of different data to provide a general view or to compute the background of a meteor shower is generally less accurate than the case of an encounter with a single trail.

As a consequence here is the choice made to define the different labels: 
\begin{itemize} 
\item "G": good quality: the forecasting is provided for a single year, is caused by a single trail ejected by an observed passage of a parent body. Typical case is the Leonids 2001, see sec. \ref{sec:appli}.
\item "F": fair: all cases that are neither good nor "poor".
\item "P": poor: the forecasting was performed using the concatenation of several years (background) for a poorly observed parent body, or for highly perturbed trails for which the close encounters happened before/after the first/last observation of the parent body.
\end{itemize}

\section{Applications to famous showers} \label{sec:appli}

\begin{figure}[!ht]
\begin{minipage}[c]{.48\linewidth}
\includegraphics[width=0.95\textwidth]{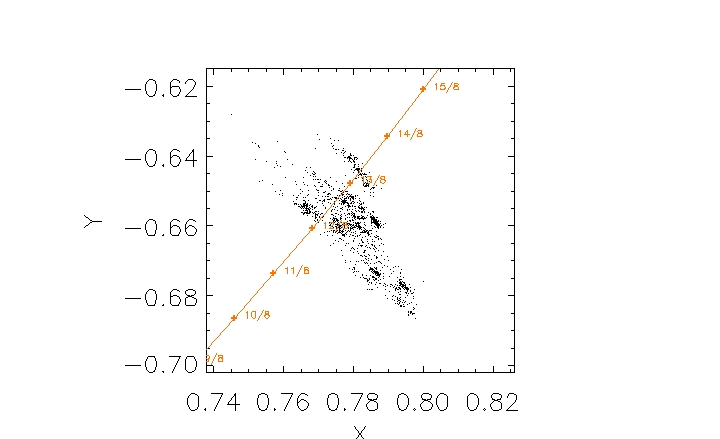}
\caption{2009 Perseids: the whole simulated stream.}
\label{Fig:PER2009}
\end{minipage}\hfill
\begin{minipage}[c]{.48\linewidth}
\includegraphics[width=0.95\textwidth]{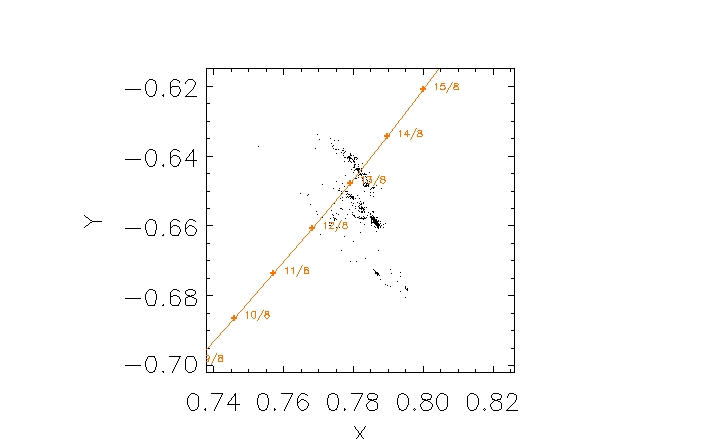}
\caption{2009 Perseids: the 441 trail.}
\label{Fig:PER2009-441}
\end{minipage}
\end{figure}

The goal of this section is to provide the reader with some direct applications and illustrations of what was briefly presented above.
For the purpose of this paper I will focus on a few famous showers.
Table \ref{tab:appli} includes several post-predictions of these showers, and the comments for each case are as follows.
\begin{enumerate}[label=Case \arabic*]
\item This famous Leonid outburst was correctly predicted. The stability of the parent orbit allowed a great confidence, enhanced by the fact that the comet 55P was observed before and after the simulated trail (1767). Although its activity at this time was not observed it worked very well.
\item Global prediction for the 2001 Leonids: the level is clearly too high, because of the contribution of several trails causing several outbursts. In such a case the predictions are not accurate at all. This method should therefore be used only for years that do not present any outburst.
\item 2003 Leonids caused by the 1499 trail: an outburst of ZHR~100/hr was predicted, but the close encounter index is very high, raising doubts regarding this number. Indeed, an outburst of only $\sim 30$/hr was reported by \cite{Arlt2004}.
\item and 4b: 2011 Draconids outburst from two different trails: one was observed (1900) but not the other (1894). The CE factor is not really different between the two cases, but the absence of observations for the 1894 trail makes the forecasting less confident. The observed level was indeed lower than expected \cite{Koten2014}.
\item 2009 Perseids from a global consideration: only %
 five perihelion returns of the comet were observed, out of 17 simulated. 
\item 2009 Perseids outburst caused by the 441 trail, by selecting all the particles crossing the path of the Earth  (i.e. ejected at all perihelion passages). The trail is highly perturbed and intersects the Earth at several different times. The level of the shower is computed at the time of maximum. Because the time of maximum is computed as an average (or median) position of the particles (defining an average/median location of the stream), considering all the particles introduces a bias, because of the spread of the location of all the particles.  This raises the uncertainties in the timing and level of the shower: see next item as well as Fig \ref{Fig:PER2009-441} and \ref{Fig:PER2009}.
\item 2009 Perseids outburst caused by the 441 trail, by selecting only the particles crossing the path of the Earth (see Fig \ref{Fig:PER2009-441}). This time, the particles look less spread. As a consequence the time of maximum better corresponds to the densest part of the stream encountered by the Earth. As a consequence, more particles are taken into account to compute the level. The computed level is slightly higher than previously derived. This case illustrates the importance of point \ref{evolv}.
\item Expected 2017 Quadrantids: the encounter factor is so high that such a prediction cannot be taken seriously.  For recent modeling of the Quadrantids see \cite{Abedin2015}. 
\end{enumerate}

\begin{sidewaystable}
\centering
\caption{Forecasting of past and future meteor showers. Comments for each line are provided in text with reference to the Id. ZHR is from the model (not observed) and the unit is $hr^{-1}$.}             \label{tab:appli}      
\begin{tabular}{l c c c c c c c r}        
\hline\hline                 
Id & Planet & Shower & parent  & year & trail(s) & ZHR & conf. index  & Quality label \\    
\hline                        
1 & Earth         &        Leonids       &     55P & 2001   & 1767          &  1669    &    SYO0/1CE0.00 & G \\
2 & Earth         &        Leonids       &     55P & 2001   & 604 to 1998  &  1500    &   GYO4/43CU0.39 & F\\
3 & Earth         &        Leonids       &     55P & 2003   & 1499          &  100       &   SYO0/1CE10.11 & F \\
\hline
4 & Earth        &        Draconids    &      21P & 2011 &   1900  &   214  &     SYO1/1CE0.00 & G\\
4b & Earth        &        Draconids    &      21P & 2011 &   1894  &   141  &     SYO0/1CE0.90 & F\\
\hline
5 & Earth        &        Perseids       &      109P & 2009 &   -68 to 1992  &   159.  &     GYO5/17CU0.00 & F\\
6 & Earth        &        Perseids       &      109P & 2009 &    441   &    35.    &    SYO0/1CE0.00 & F\\
7 & Earth        &        Perseids       &      109P & 2009 &    441  	&    40.    &    SYO0/1CE0.00 & G\\
\hline                                   
8 &   Earth      &  Quadrantids       &  2003EH1 & 2017   &   1704 to  2003  &     7.    &  GYO1/57CU1566 & P\\
\hline                                   
\end{tabular}
\end{sidewaystable}

\section{Conclusion} \label{sec:conc}

Several new methods to perform the forecasting, by concatenating some data are entirely new, and are inherent to the very method used (in my case \cite{Vaubaillon.et.al2005}).
This meteor shower level forecasting confidence index is, to my knowledge, the first attempt to provide the community with a way to better understand how such tasks are performed in a sufficient way that the end user has an idea of how much (s)he can trust the results.
In such, this is not perfect, and a thorough calculation might be necessary in the future, though at this point it seems an overkill to me.

An effort of simplification of this index was recently requested to the author, but additional work is needed at this point in order to keep a way to provide a lot of information in a concise way.
This was my wish when constructing this index, and I hope future works will be able to bring improvements without taking anything from it.

\section*{Acknowledgements}
I am especially thankful to G. Valsecchi for his insight and help regarding the quantification of the effect of close encounters with the planets, as well as to D. Asher and the anonymous referee who helped making this paper better. All numerical simulations were performed at CINES supercomputer facility, Montpelier, France.



 \bibliographystyle{elsarticle-harv} 
 \bibliography{bibtex}





\end{document}